\documentclass[epj]{webofc}
\usepackage[utf8]{inputenc}
\usepackage[varg]{txfonts}   
\usepackage{booktabs}
\usepackage{xcolor}
\definecolor{darkred}{rgb}{0.4,0.0,0.0}
\definecolor{darkgreen}{rgb}{0.0,0.4,0.0}
\definecolor{darkblue}{rgb}{0.0,0.0,0.4}
\usepackage[bookmarks,linktocpage,colorlinks,
    linkcolor = darkred,
    urlcolor  = darkblue,
    citecolor = darkgreen]{hyperref}
\newcommand{\beq}{\begin{equation}}
\newcommand{\eeq}{\end{equation}}
\newcommand{\bea}{\begin{eqnarray}}
\newcommand{\eea}{\end{eqnarray}}

\newcommand{\nn}{\nonumber}
\newcommand{\e}{\rm e\,}
\newcommand{\tr}{\hbox{tr}}

\newcommand{\Det}{{\cal D}}
\newcommand{\im}{\hbox{i}}
\newcommand{\ims}{\hbox{\footnotesize i}}
\newcommand{\Rep}{\hbox{Re}}
\newcommand{\Imp}{\hbox{Im}}
\newcommand{\half}{\hbox{\text {$\frac{1}{2}$}}}
\newcommand{\third}{\hbox{\text {$\frac{1}{3}$}}}
\newcommand{\diag}{\hbox{\text {diag}}}

\newcommand{\C}{\mathbb{C}}

\newcommand{\R}{\mathbb{R}}

\usepackage{subfigure}
\wocname{EPJ Web of Conferences}
\woctitle{Lattice2017}
\begin{document}
%
\selectlanguage{english}
\title{%
Getting even with CLE
}
\author{%
\firstname{Gert} \lastname{Aarts}\inst{1}\fnsep\thanks{Acknowledges 
financial support from STFC, the Royal Society and the Wolfson Foundation.} 
\and
\firstname{Kirill} \lastname{Boguslavski}\inst{2}\fnsep\thanks{Acknowledges
finacial support from the European Research Council under grant 
No.~ERC-2015-COG-681707.} 
\and
\firstname{Manuel}  \lastname{Scherzer}\inst{3}
\fnsep\thanks{Acknowledges financial support  from DFG under under 
grant STA 283/16-2.}  
\and
\firstname{Erhard} \lastname{Seiler}\inst{4}
\fnsep\thanks{Acknowledges financial support  from DFG under under 
grant STA 283/16-2.} 
\and
\firstname{D\'enes} \lastname{Sexty}\inst{5}
\fnsep\thanks{Acknowledges financial support from DFG Heisenberg 
fellowship under grant 
SE 2466/1-1.} 
\and
\firstname{Ion-Olimpiu}  \lastname{Stamatescu}\inst{3}\fnsep\thanks{Speaker, 
\email{stamates@thphys.uni-heidelberg.de}, acknowledges 
financial 
support from DFG, grant STA 283/16-2.}
}
\institute{%
Physics Dept., Swansea Univ., UK, 
\and
Physics Dept., Univ. of Jyv\"askyl\"a, Finland, 
\and
Institute for Theoretical Physics, Univ. Heidelberg, Germany, 
\and
Max Plank Institute for Physics, M\"unchen, Germany, 
\and
Physics Dept., Wuppertal Univ, Germany, 
}
\abstract{%
  In the landscape of approaches toward the simulation of 
Lattice Models with complex action the Complex Langevin (CL) 
appears as a straightforward method with a simple, well defined setup. 
Its applicability, however, is controlled by certain specific conditions 
which are not always satisfied. We here discuss the 
procedures to 
meet these conditions and the estimation of  
systematic errors and present some actual achievements.
}
\maketitle
\section{Introduction}\label{intro}

For a model defined by a path integral with a complex action { $S$},
CL is a stochastic process proceeding on the complexification 
of the original manifold, e.g., ${\R}^n \longrightarrow
\C^n$
   or $SU(n) \longrightarrow SL(n,\C)$. It
   involves  a drift term { $K = -\nabla S$ }
and 
a suitably normalized random noise {$\eta$}
\cite{review}, cf also \cite{Seiler:thiscontrib65}.
 E.g., for one variable 
 { $x \rightarrow z = x + \im \, y$ } the
 CL amounts to two real 
 Langevin processes in the process time {  $t$}
 \bea
 {\delta x(t)} &=& K_x(z)\,\delta t + 
\eta_x (t)\, , \quad K_x=\Rep K(z),\quad
  \langle \eta_x \rangle= 0\,,  
  \quad\langle \eta_x^2\rangle =
 2 N_R\,\delta t\, \nn\\
  {\delta y(t)} &=& K_y(z)\,\delta t + 
\eta_y (t)\, , \quad K_y=\Imp K(z),\quad
  \langle \eta_y \rangle= 0\,,  
  \quad\langle \eta_y^2\rangle =
 2N_I\,\delta t\, , \quad N_R - N_I =1 \, .\nn
\eea
The process realises a real probability distribution { $P(x,y;t) $}
accompanying the complex distribution  $\rho(x,t)$ (with
asymptotic limit { $\rho(x) =\exp(-S(x))$}). They   
evolve according to  real (complex) FPE's:
\bea
&&\partial_t  P(x,y,t) = L^T P(x,y,t)\, , \quad
L =
(N_R\partial_x + K_x(z))\partial_x +   
(N_I\partial_y + K_y(z)\partial_y \nn\\
&&\partial_t \rho(x,t) = L^T_0 \rho(x,t)\,, \quad  L_0 = 
(\partial_x + K(x))\partial_x  . \nn
\eea
Notice that an associated Random Walk process can also be defined:
\vspace{-0.1cm}
 \bea
 \hspace{-1cm}&&{\delta x(t)} =\ 0 \ \ \hbox{\small{with\,pbb.}}\ \ 1-q, \quad \rm{or}\ \ 
 \pm \omega_x \ \ 
 \hbox{\small{with\,pbb.}}\ \ \half q\, (1\pm \tanh (\half \,\omega_x K_x /
 \sqrt{N_R})) ,\quad 
 \omega_x = \sqrt{2 \delta t /q} \nn\\
 \hspace{-1cm}&&{\delta y(t)} =\ 0 \ \ \hbox{\small{with\,pbb.}}\ \ 1-q, \quad \rm{or}\ \ 
 \pm \omega_y \ \ 
 \hbox{\small{with\,pbb.}}\ \ \half q\, (1\pm \tanh (\half \,\omega_y K_y /
 \sqrt{N_I})) ,\quad 
 \omega_y = \sqrt{2 \delta t /q} .\nn
\eea
The parameters $q,\, N_R,\, N_I$ can be chosen adequately; typically 
one takes $N_R=1,\,N_I=0$ 
and $q \simeq 1$.

\section{Conditions for convergence}\label{cond-conv}
For analytic observables {$O(z)$} one can formally prove \cite{Aarts:2009uq},
\cite{Aarts:2011ax}
\bea \quad
 \int {O}(z) P(x,y;t)  dxdy = \int  {O}(x) \rho(x;t) dx  . \label{e.fund}
\eea
 The proof of convergence relies on the holomorphy of the action and
 thus of the drift { $K$}, and on a well behaved distribution $P$ at large
 values of the variables. Non-holomorphic behaviour is typically 
 associated with zeroes of the measure  $\rho$ leading to a meromorphic
 drift { $K$}. Holomorphy can be regained by cutting small regions around the 
 poles. The problem reduces therefore to the control of boundary terms both
  at $\infty$ and on the small contours
around the poles. Their vanishing  
would ensure 
{ \bea
\int \left(L^T P(x,y;t-\tau)\right) O(x+ \im y;\tau)\, dxdy 
 - \int  P(x,y;t-\tau) L\,O(x+\im y;\tau) \, dxdy = 0;
\label{interpol}
\eea }
for any $\tau$ using integration by parts and thus Eq. (1)   
\cite{Aarts:2017vrv}, \cite{Nagata:2016vkn}, cf. also \cite{Seiler:thiscontrib65}.

 In practice these conditions require quantitative estimations of the behaviour
   of $P(x,y)$ at large $y$  or near the poles of the drift $K(x,y)$  
   \cite{Aarts:2017vrv}, which in realistic cases need on-line or 
   a posteriori estimations. This implies approximations
   in interpreting the data and induces systematic errors. This is 
   familiar from other simulations algorithms: reweighting, R-algorithm,
   cooling or Wilson flow,
 etc. 
   
The CL process can be redefined in many ways while still leading to the
same expectation values, see discussion in \cite{Aarts:2012ft} and further
work cited there. In particular for gauge theories one efficient method 
is to keep the process near the $SU(3)$ manifold by gauge transformations 
({\em gauge cooling} \cite{Seiler:2012wz}) minimising some measure
of non-unitarity, such as a ``unitarity norm'' { ${\cal N}_U =
\tr(U\, U^{\dagger})-3$},
$U \in SL(3,\C)$. 
One can envisage various 
ways of defining such cooling procedures by minimising also other 
quantities, see e.g. 
\cite{Nagata:2016vkn}. 
Other
transformations of the process have been investigated
\cite{Attanasio:2016mhc}.

\section{Lessons from simple models}\label{simple}

We shall consider here an effective model defined as
the nontrivial link { $U \in SL(3,\C)$} of a 
temporal gauge Polyakov loop in the field of its neighbours. 
Further simple
models are discussed in  \cite{Seiler:thiscontrib65},
\cite{Aarts:2017vrv}.
After transforming to the 
Cartan basis the action reads:
{\bea
-S ={\beta} \sum_{i=1}^3 \left( \e^{\alpha_i}\e^{\ims w_i} +
\e^{-\alpha_i} \e^{-\ims w_i} \right) + \ln \Det +  \ln H , \quad
\sum_{i=1}^3 w_i =0, \  U =\diag\{\e^{\ims \,w_i}\} ,\  O_n =\tr U^n
\label{e.effobs}
\eea
}
with $H$ the $SL(3,\C)$ Haar measure. $\alpha \in \C$ simulate the staples. 
The fermionic determinant is:
{\bea
\hspace{-0.5cm} &&{\cal D} = \left( D\,{\tilde D}\right)^2 , \ 
D =  
\left(1+C^3\right)\left(1+a\,P + b\,{\tilde P}\right), \quad
{\tilde D} =  
\left(1+{\tilde C}^3\right)\left(1+{\tilde a}\,{\tilde P} + 
{\tilde b}\,P\right),
 \label{e.effdet1} \\
\hspace{-0.5cm} &&C=(2\,\kappa \e^{\mu})^{N_\tau},\
 a= 3 C/(1+C^3),\  b= C\,a,  \quad \quad {\tilde C}=(2\,\kappa \e^{-\mu})^{N_\tau},
 {\tilde a}= 3 {\tilde C}/(1+{\tilde C}^3),\  {\tilde b}= 
 {\tilde C}\,{\tilde a} \nn\\
 \hspace{-0.5cm} && P = \third \tr U ,\quad  {\tilde P} = \third \tr U^{-1}
\eea
At large $\mu$ we can set ${\tilde D} \simeq 1$. 
The parameters $a,\, b$ have maxima of height $2^{2/3}$ at 
$C=2^{\mp 1/3}$.  We expect effects from the
zeroes of the determinant 
when $a,\, b >1$. 
We present 
 results in terms of $\mu$ for a targeted 4-dim.
 lattice model with $N_{\tau} =8$ 
 and $\kappa=0.12$. The only relevant parameter is $C$
  and the interesting region is around $C=1$ which means
$\mu \simeq 1.425$. The ``bad'' interval $a,\,b>1$ corresponds to
$1.29 <\mu<1.56$. Larger $\kappa$ shifts the 
$a,\,b$ peaks
 to smaller $\mu$, 
larger $N_{\tau}$ makes them sharper. 

To illustrate the effect 
of the neighbours we consider two cases (``ordered'' and ``disordered''):

 \hspace{0.5cm} 0:\, $\{\alpha\}=0$, \quad \quad
1:\, $\{\alpha\} =\{ 0.2+1.5\,\im;\, -0.2+3.1\,\im;\, 0.2-0.7\,\im\}.$

\begin{figure}[h] 
  \centering
  \includegraphics[width=6.2cm,clip]{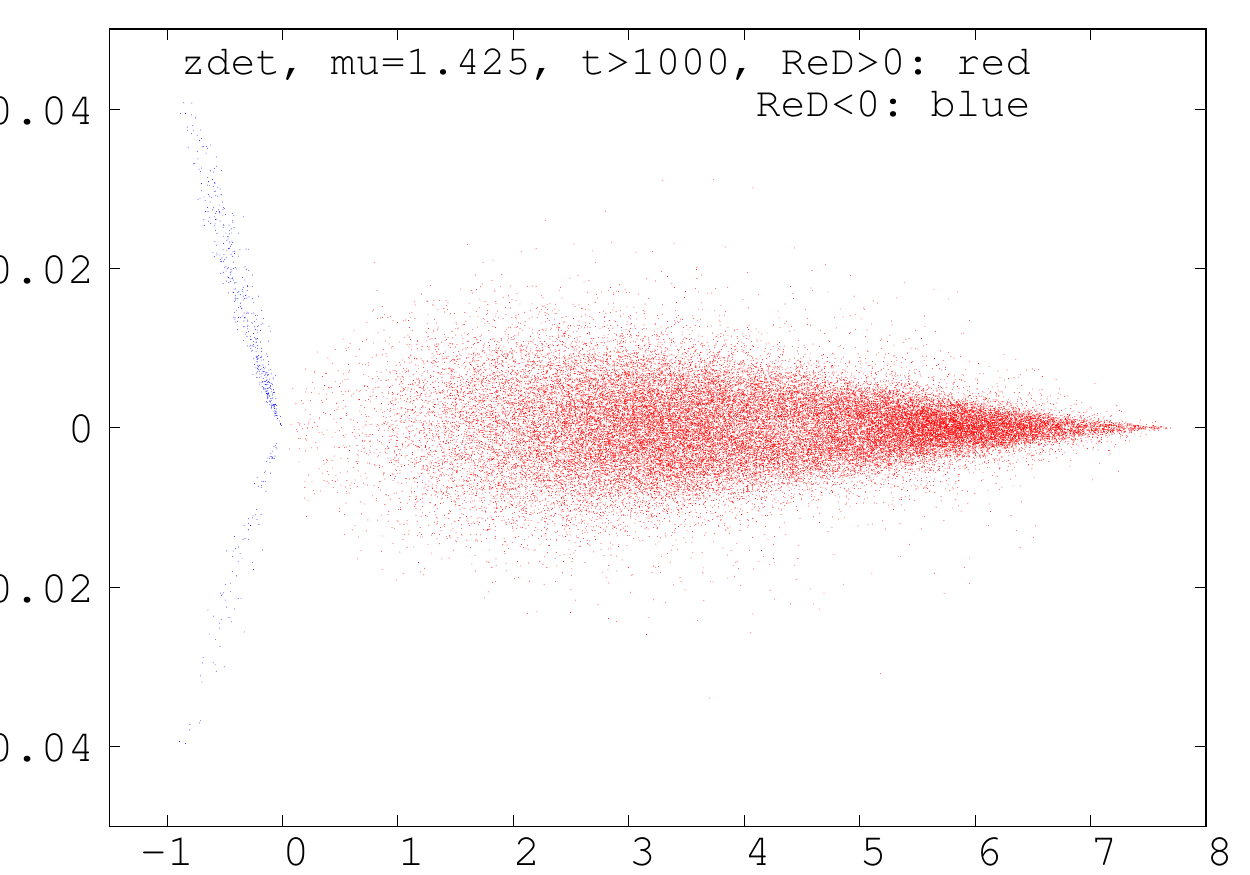}
  \includegraphics[width=6.2cm,clip]{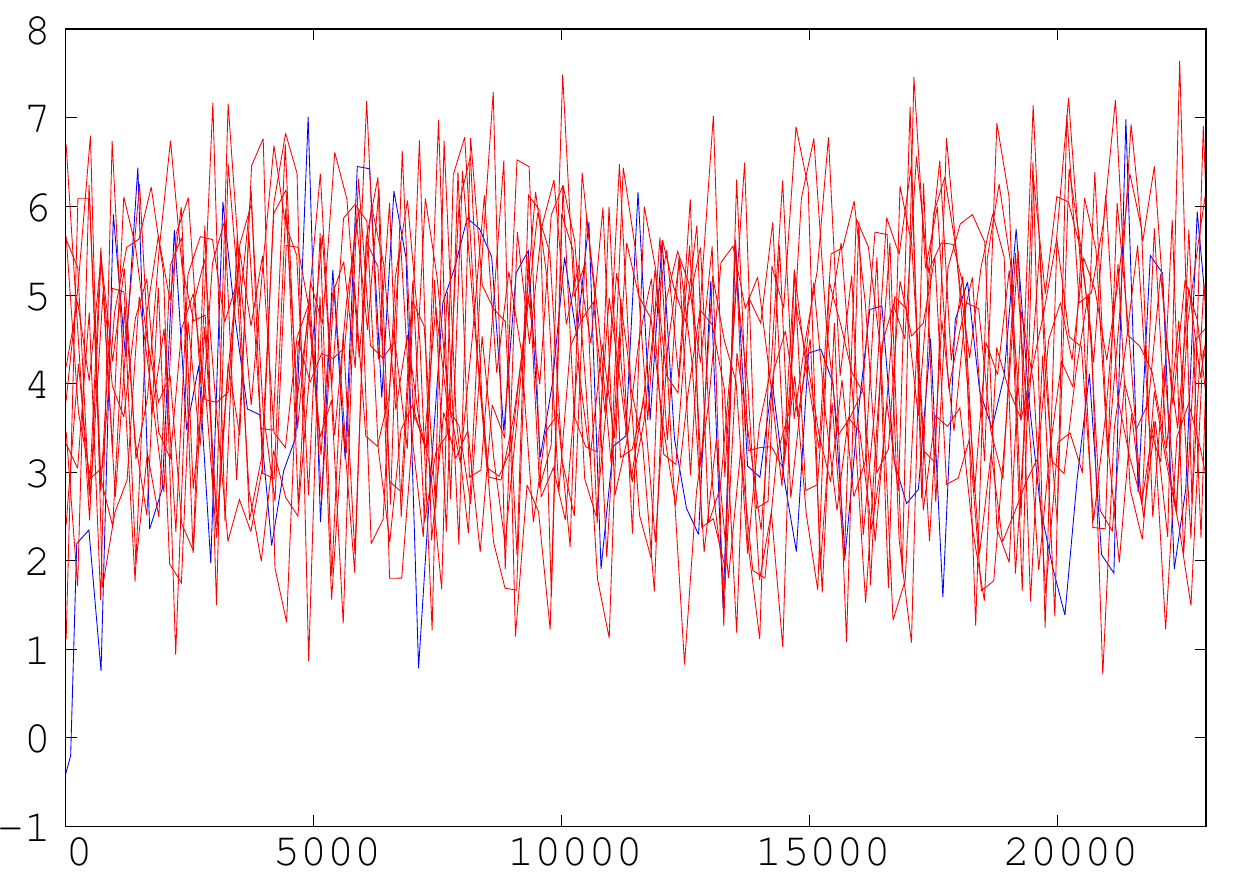}
  \includegraphics[width=6.2cm,clip]{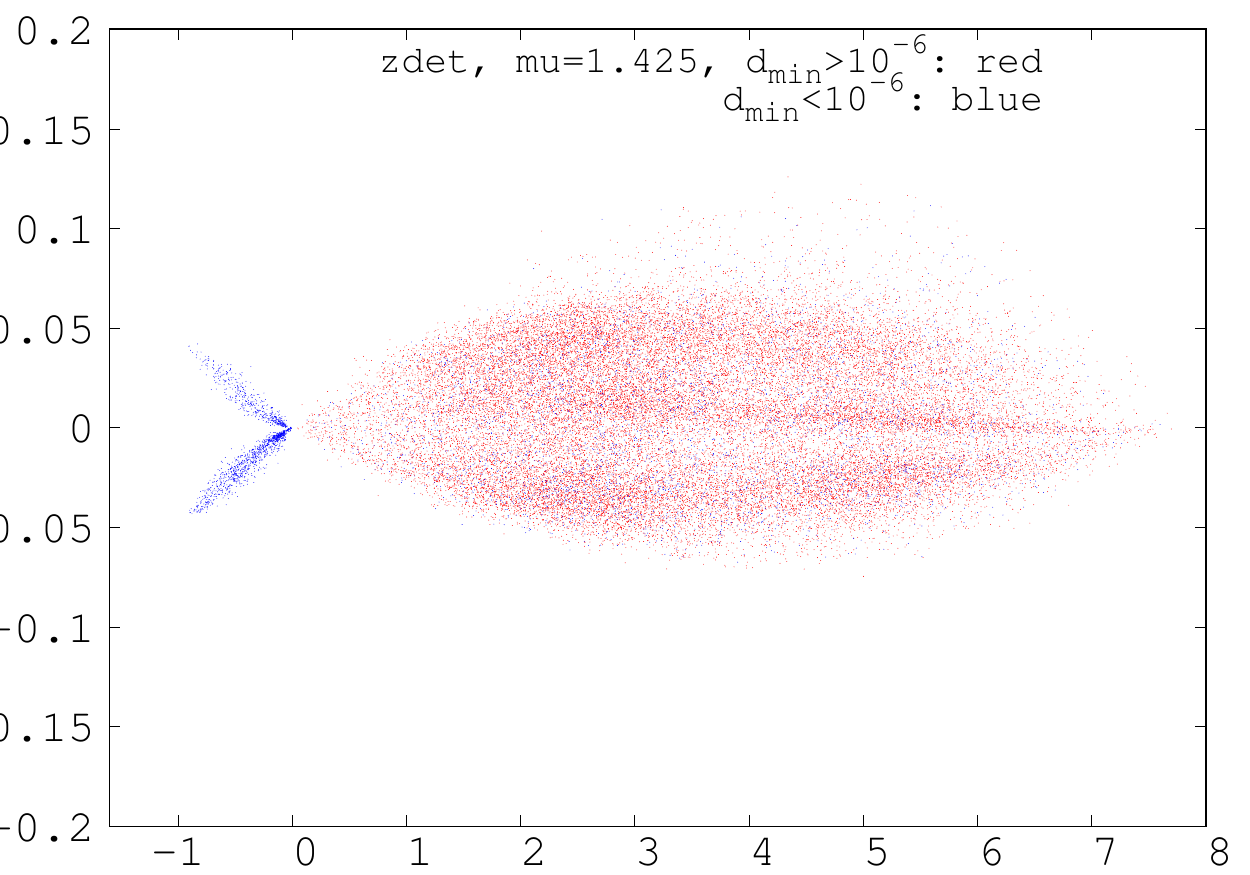}
  \includegraphics[width=6.2cm,clip]{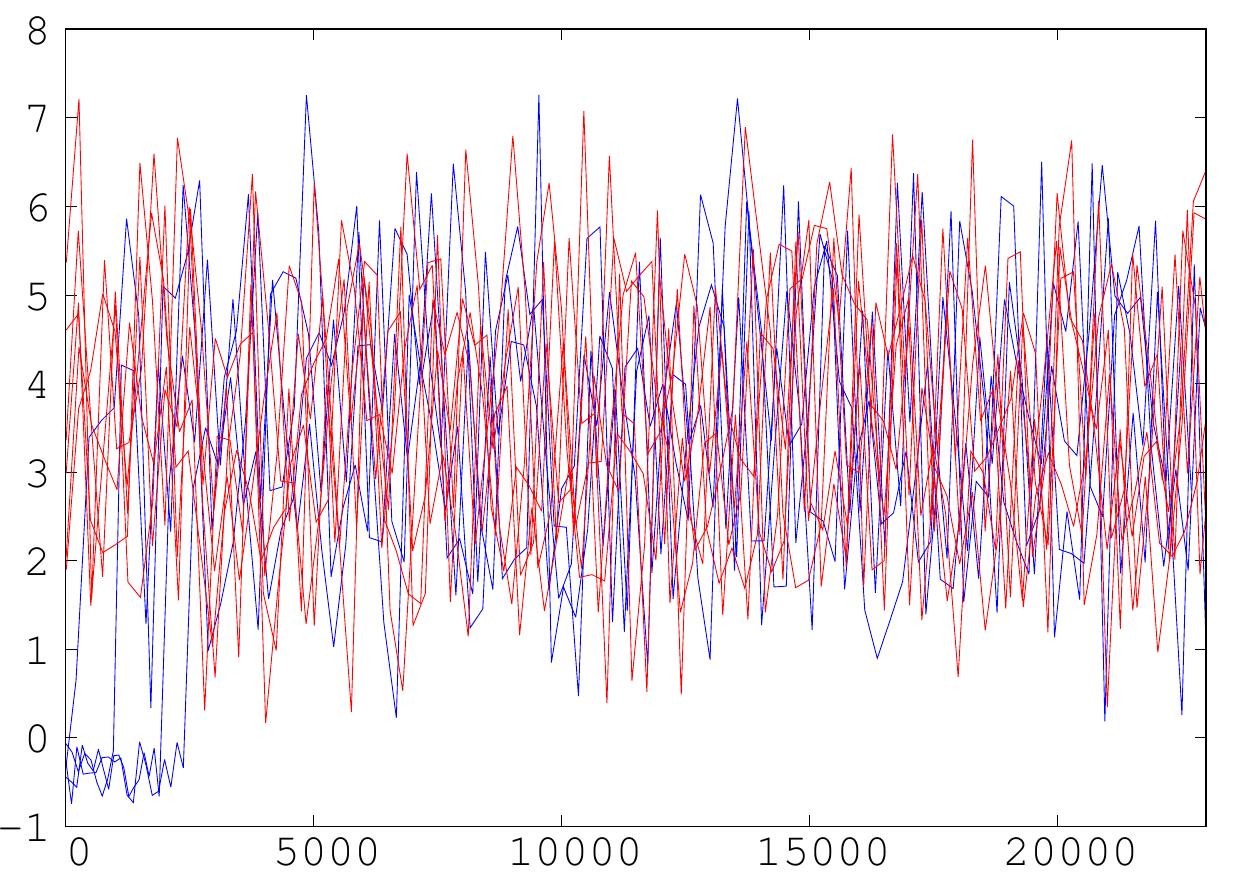}
  \includegraphics[width=6.2cm,clip]{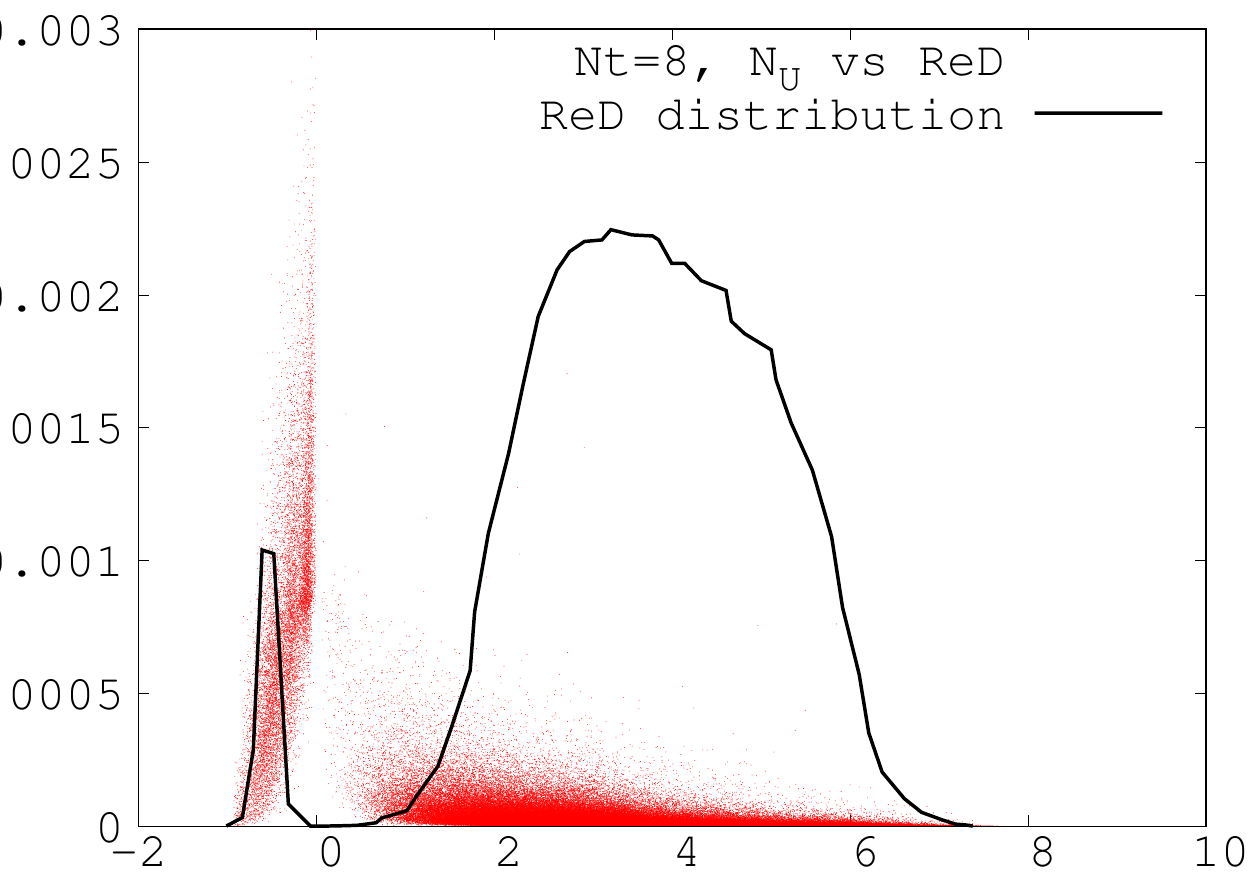}
  \includegraphics[width=6.2cm,clip]{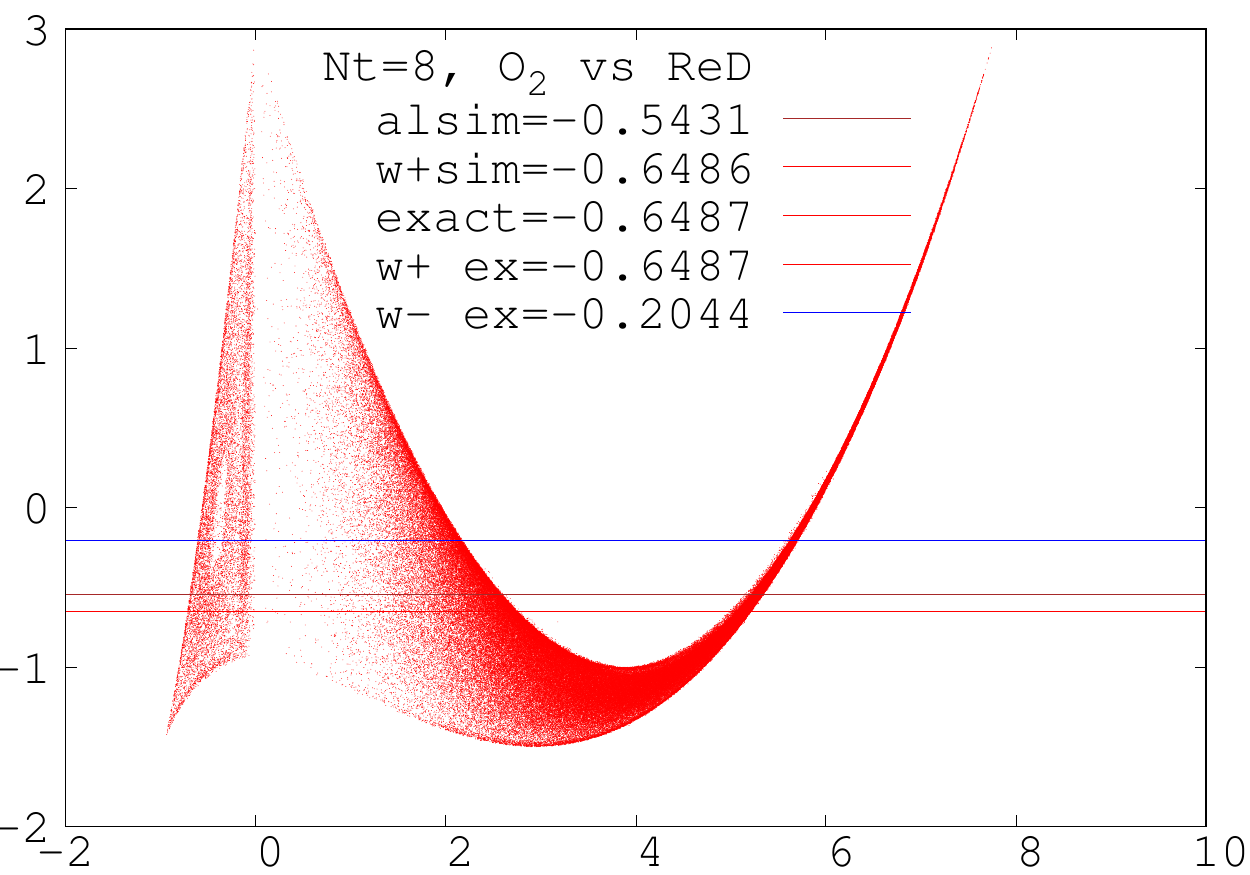}
  \caption{One link effective model,
  $N_{\tau} =8,\, 
  \kappa=0.12$,  $\mu=1.425$ (``bad'' point). 
 {\it Top}: scatter plot of $D$ 
 and trajectory histories for the ordered case. {\it Middle}: same for the
 disordered case.
 {\it Bottom}:  histogram of $\Rep D$ and 
 correlation of  ${\cal N}_U$ and $\Rep D$ (left plot)  and  of 
 the observable $O_2$
 and $\Rep D$  (right plot). See also the text.}
\end{figure}
The scatter plots for $D$ show a peculiar structure in 
 the ``bad" interval (absent 
 outside of it). 
 Here the process also samples two exceptional  
 regions (blue in the upper plots of Fig. 
 1)
 characterized by $\Rep D <0$. They 
 contribute wrong values for the observables. They are
 connected by a bottleneck at $D=0$ to the ``red" region, therefore 
 $\Rep D <0$ contributions come from trajectories which went near
 $D=0$.
Trajectories  starting in ``blue" 
 typically switch to ``red" and only rarely visit 
 ``blue'' again.

 From the relative weight of the ``blue" regions
  one estimates a systematic error of about $1/1000$,
   in agreement
  with observed deviations - see the histogram on left bottom plot of Fig. 1. 
   Discarding the ``blue'' 
 contributions leads to 
  good
   results 
 in the ordered case, see Fig. 1 right bottom plot comparing the simulation 
 and the exact results from all and from the ``red''/``blue'' ($w^{\pm}$) regions. 
 See Fig. 2 
 for an extended comparison in the ordered case 
 (the agreement is less good in the disordered case). 

\begin{figure}[thb] 
  \centering
  \includegraphics[width=6.2cm,clip]{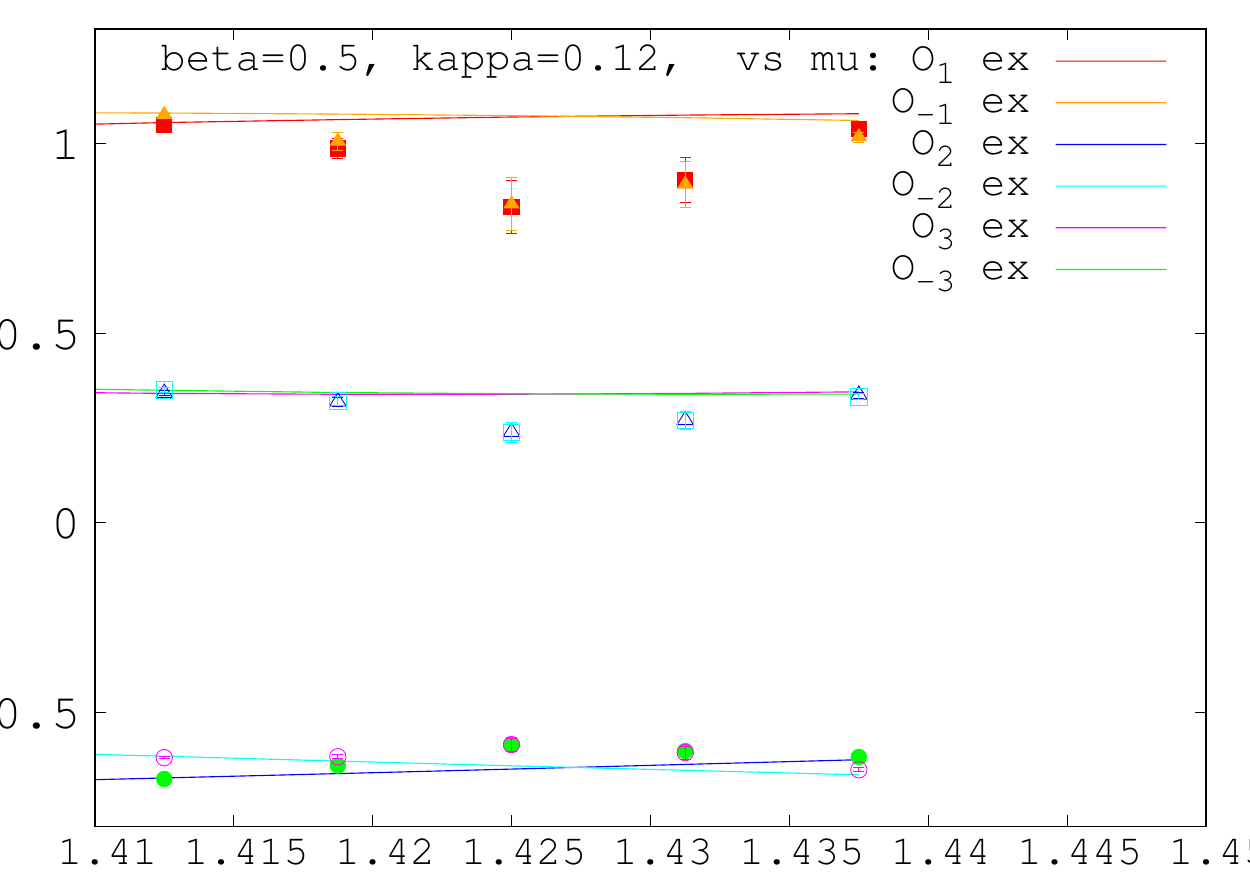}
  \includegraphics[width=6.2cm,clip]{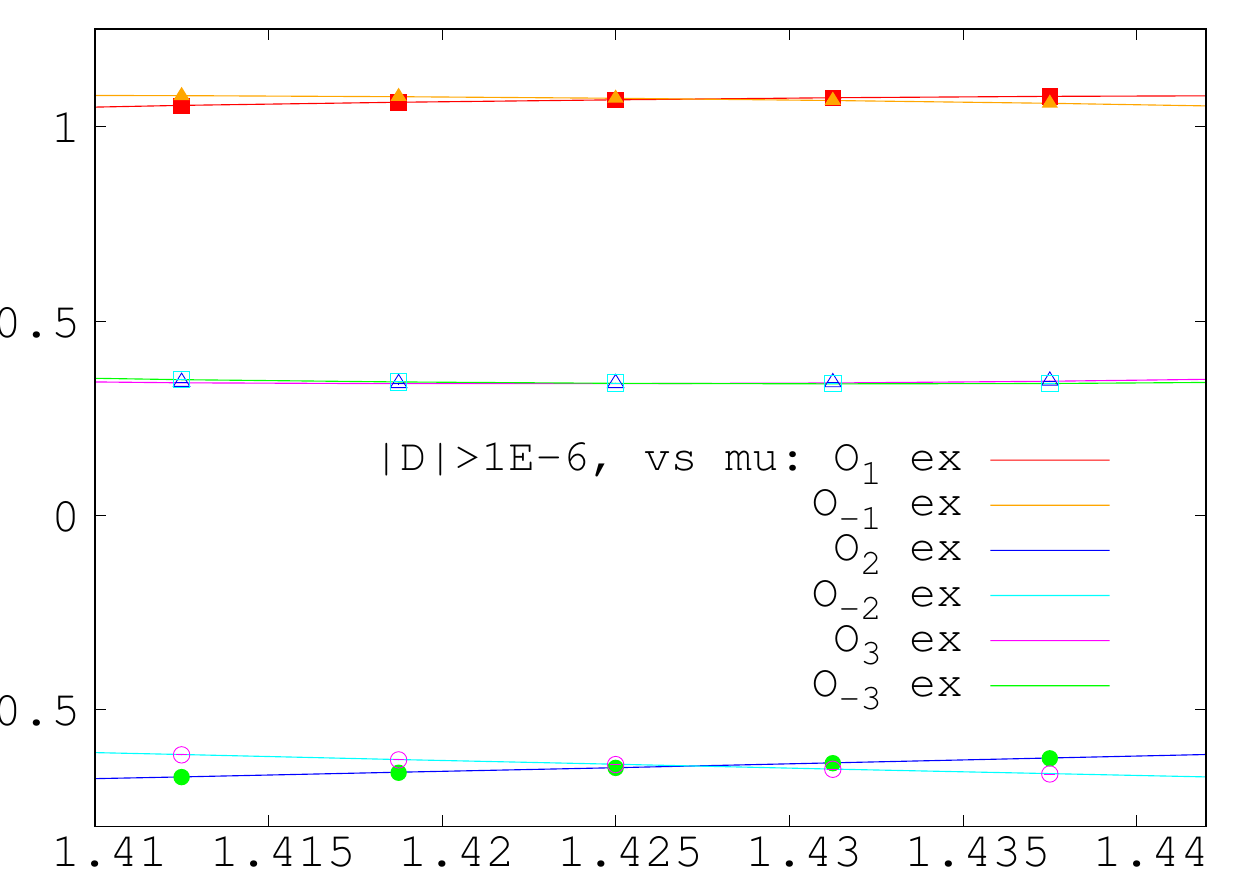}
  \caption{
  One link  model,
  $N_{\tau} =8,\, 
  \kappa_{N_{\tau}}=0.12$: Observables vs $\mu$ from all
  regions and from the ``red'' region.}
  \label{fig-2}
\end{figure}

\section{From simple models to QCD}\label{simple-QCD}

\subsection{The Polyakov chain model}\label{pol}
A Polyakov chain of $N_{\tau}$ links allows to test procedures involving 
 full gauge field integration of many links \cite{Seiler:2012wz}. 
Since it can be reduced by gauge transformations
to the one link model of Sec. \ref{simple} one can compare with exact results. 
On the other hand, the direct integration in the full group space over the 
$N_{\tau}$ links
allows to see the effects
of accumulation of round-off errors, cooling, etc. This model is a step
toward HD-QCD which is no longer exactly solvable.

We shall consider two versions, referred to as model 1 and model 2 below:
\bea
1:\ -S_1&=&\beta(P + {\tilde P}) + 2 n_f \ln(1+a P + b {\tilde P}), \quad
2:\ -S_2 =(\beta + 2 n_f a)P + (\beta + 2 n_f b){\tilde P}, \label{e.s2}\\ 
P &=& \tr \prod_{t=1}^{N_{\tau}} U_t, \quad \quad {\tilde P} = 
\tr \left(\prod_{t=1}^{N_{\tau}}
 U_t \right)^{-1} 
\label{e.p}
\eea
with $a,\, b$ of Sec. \ref{simple} and $n_f$ the number of flavours. 
CL proceeds in the group algebra:
\bea
U'_t = \exp \left ( -\sum_a \im\,\lambda^a (\epsilon D^a_t S[U] + \sqrt{\epsilon}
\eta_t^a )\right) U_t 
\eea
with $D^a$: covariant derivative.
Model 1 is the direct extension of that in Sec. 3,
 model 2 is the un-resummed version of model 1 \cite{Stamatescu:2016dwx} and has no zeroes.
We test  two ``cooling" methods to
control the drift in the noncompact direction by  minimising:
\bea
{\rm I:}\quad {\cal N}_U = \sum_t (\tr(U_t\, U_t^{\dagger})-3 ), \quad \quad \quad
{\rm II:}\quad {\cal N}_I = \sum_t \sum_a \left(\Imp D^a_t S[U]\right)^2
\eea
This proceeds by non-compact gauge transformations using the covariant 
gradients of the norms. In Table \ref{tab-1} we show typical results for 
$\beta=0.5,\,N_{\tau}=8,\,
\kappa=0.12,\, \mu= 1.4,\, \epsilon=10^{-5}$(non-adaptive).
We cool after each updating step:
 one sweep Cooling I (1 0),\ one sweep Cooling II (0 1), or
\ both (1 1).\\ 
The best results are obtained with more cooling sweeps. Remarks:\\
- It appears important to start in or near the
$SU(3)$ manifold and to cool during thermalisation;\\
- The value of ${\cal N}_U$ can vary (if it stays below $~0.1$), especially with (1,1) 
cooling;\\
- For model 1 cooling also seems to improve concerning the effects of the 
non-holomorphy.
\begin{table}[thb]
  \small 
  \centering
  \caption{Comparing cooling types for the Polyakov chain model.
  The statistical errors are  $O(10^{-3})$.}
  \label{tab-1}
  \begin{tabular}{llllllll}\toprule
  $P$ exact:	&model 1:&$P_{full}=1.2775$&$(ph.q.: 1.2873)$	&&  model 2:&$P_{full}=1.5637$&$(ph.q.: 1.5822)$\\\midrule
  cooling	&${\cal N}_U$	&${\cal N}_I$	&$P$	&&${\cal N}_U$	&${\cal N}_I$	&$P$\\\midrule
  1\ 0	&$0.56\,10^{-4}$&$0.17\,10^{-5}$&$1.2773$&&$0.22\,10^{-2}$&$0.89\,10^{-5}$&$1.5676$\\																												
  0\ 1	&$0.89\,10^{-3}$&$0.15\,10^{-5}$&$1.2800$&&$0.13\,10^{-1}$&$0.95\,10^{-5}$&$1.5658$\\																												
  1\ 1	&$0.48\,10^{-4}$&$0.10\,10^{-5}$&$1.2734$&&$0.21\,10^{-2}$&$0.85\,10^{-5}$&$1.5695$\\ \bottomrule
  \end{tabular}
\end{table}

\subsection{HD-QCD}\label{HD}
HD-QCD arises in the limit $\kappa \rightarrow 0,\ \mu \rightarrow \infty$ 
where
the
determinant of QCD becomes a product of factors Eq. (4)
over all Polyakov loops \cite{Bender:1992gn}. Calculations have explored
among other things
the phase diagram of the model , see e.g. \cite{DePietri:2007ak} (reweighting), 
\cite{Aarts:2016qrv} (CL).

As before, the correctness of the results 
depends on the weight of the $\Rep D <0$ regions, which was there typically
$O(10^{-2}-10^{-4})$ and we found a corresponding systematic error of
the same magnitude.
Now this weight and the associated systematic error are difficult to estimate, 
due to the superposition of
single regions in the product of factors which define the determinant. 
 We 
therefore need
direct simulations and tests. 
Comparing CLE and reweighting in the region where
the latter is reliable ($\mu < 1$ with phase rewighting) \cite{DePietri:2007ak} 
we have observed generally good agreement but
increasing deviations for $\beta \leq 5.7$ \cite{Seiler:2012wz}. 
Lower temperatures 
require therefore large $N_{\tau}$ which would permit to work at large $\beta$. 
This appears to be a question of computer power but further tests
may be necessary.

\subsection{Full QCD in the $\kappa$ expansion}\label{exp}

Splitting the fermionic determinant into temporal and spatial contributions 
we evaluate the first factor analytically (HD-QCD) and use a hopping 
expansion for the spatial part \cite{Aarts:2014bwa},
 \cite{Fromm:2011qi}. This was shown to
converge 
towards full QCD at sensible parameter choices already 
around order ten \cite{Aarts:2014bwa}. 

As we have learned from HD-QCD 
we need  $\beta >5.7$ for the gauge
cooling to be effective.  Here we stay with $\beta=6$. 
We present for illustration
preliminary $O(\kappa^{10})$ results on lattices of $10^3\times N_{\tau}$ at 
$\beta=6, \, \kappa=0.12$, aimed at the phase diagram of 
QCD. For an investigative study trying to
reach low temperatures we
use large $N_{\tau}$ but are  aware of the finite size effects
induced by the wrong aspect ratio. 

In the upper two plots of Fig. \ref{fig:lowT} 
we fix a rather low temperature $N_{\tau}=16$ and vary $\mu$.
It appears that we can explore both the confining and the deconfining
phase. At $\mu=0$ the gauge cooling 
ensures that the results agree with those obtained by reunitarisation,
which is equivalent to a real Langevin simulation (blue and green dots 
at $\mu=0$ on the plots). Notice that the agreement
 may be lost by choosing an inadequate
 $\beta <6$ where cooling becomes ineffective \cite{Bloch:2017jzi},
\cite{Sinclair:2016nbg}. 

In the lower two plots of Fig. \ref{fig:lowT} we vary the 
temperature $T=1/N_{\tau}$
 by increasing $N_{\tau}$ at fixed $\mu$. Starting at high 
 temperature in the deconfining phase 
 for $\mu \leq 1.1 $ 
we see the signal of the transition to the confining phase at $T \sim
0.1$ ($N_{\tau} \simeq  10$), especially perspicuous in the Polyakov loops.
At  $\mu \geq 1.8 $ we see fully saturated density at low temperature and  
slightly 
decreasing at higher $T$.

In the region $0 \leq \mu \leq  1.1$ cooling is  effective 
and  we observe the silver blaze phenomenon expected in the
confining region already at $N_{\tau} = 16$. Likewise the region of $\mu \geq 1.8$ 
is accessible  and shows the expected saturation effects. We also see from
both sides the signal of approaching a transition, but we could
not enter the region $1.1 < \mu < 1.8$ in order to resolve it
 with only standard gauge cooling. For the further developments
we might be forced  to
resort  to further procedures such as dynamical stabilisation 
\cite{Attanasio:2016mhc} and other improvements,
\cite{Schmalzbauer:2016pbg}.
This is work in progress. See also \cite{Sexty:2013ica}.
\begin{figure}[h] 
  \centering
  \includegraphics[width=6.2cm,clip]{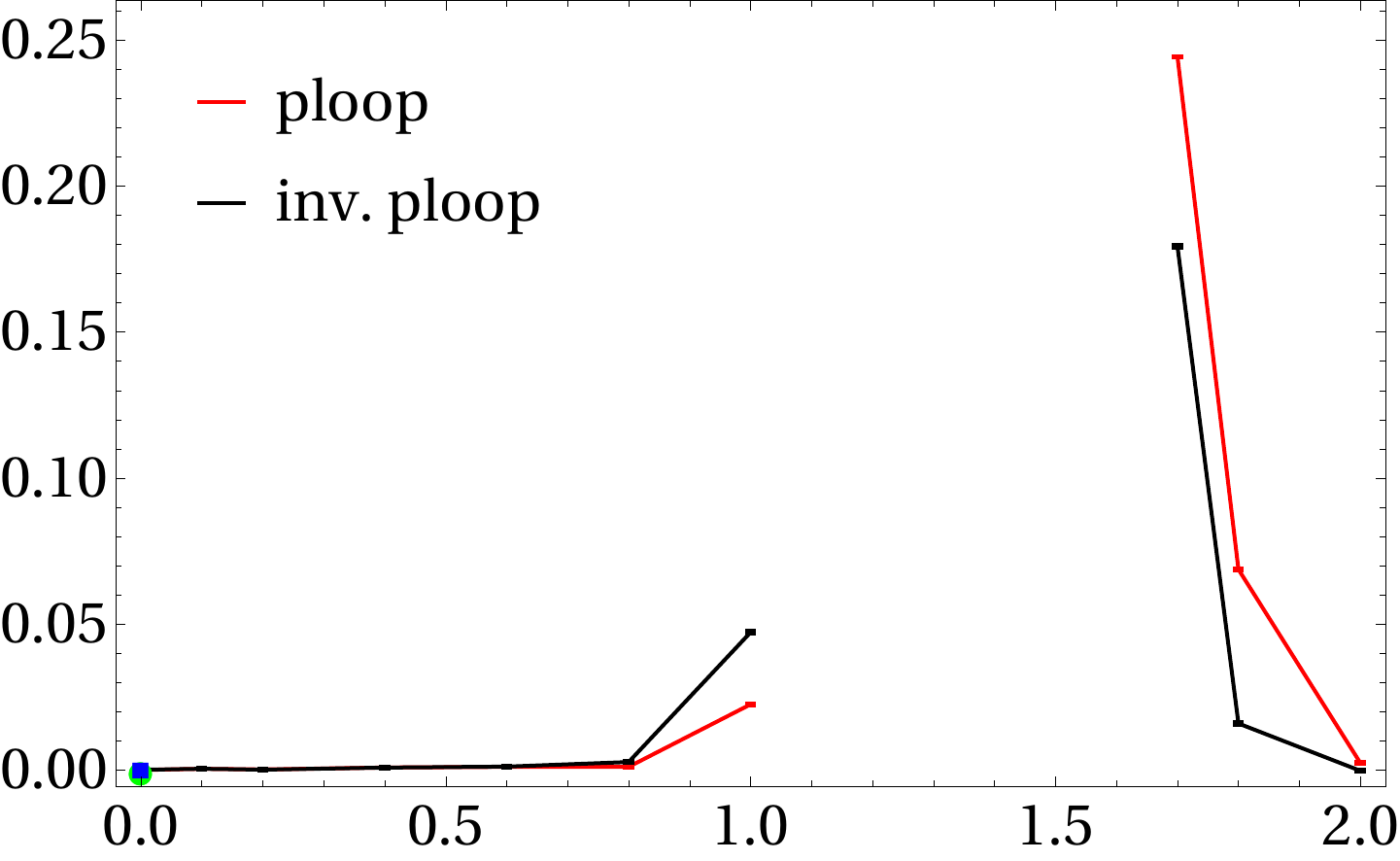}
  \includegraphics[width=6.2cm,clip]{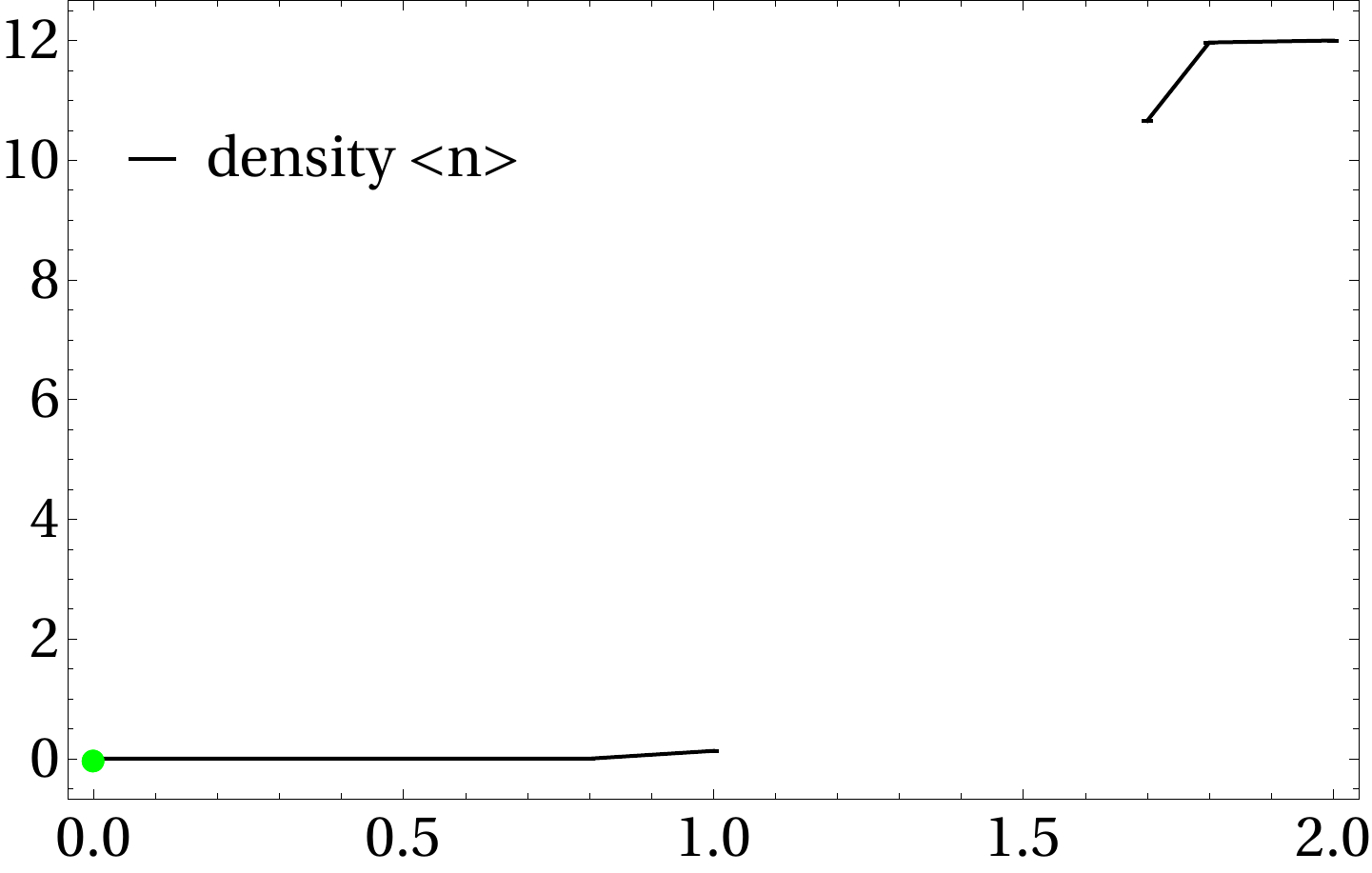}
  \includegraphics[width=6.2cm,clip]{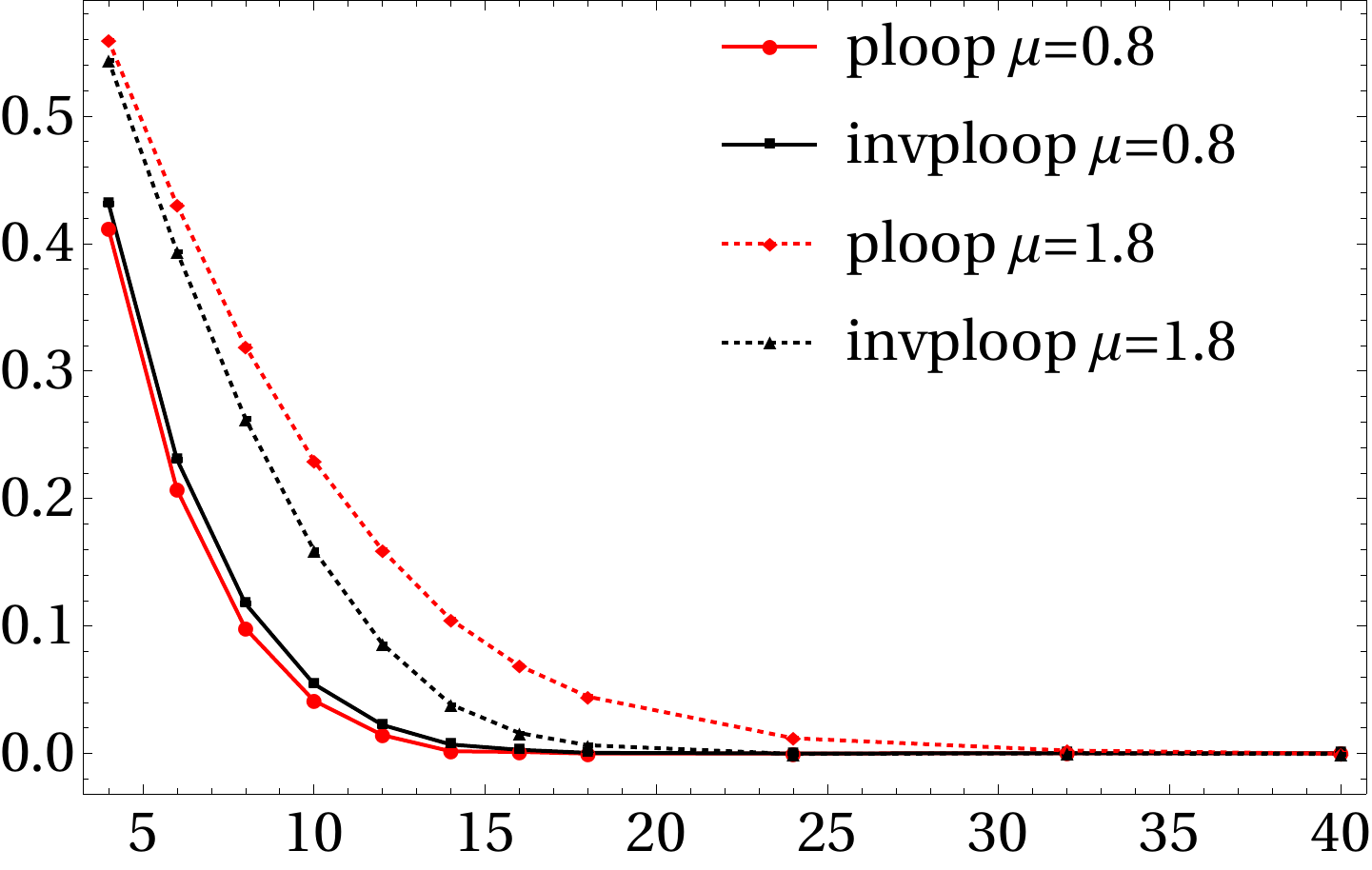}
  \includegraphics[width=6.2cm,clip]{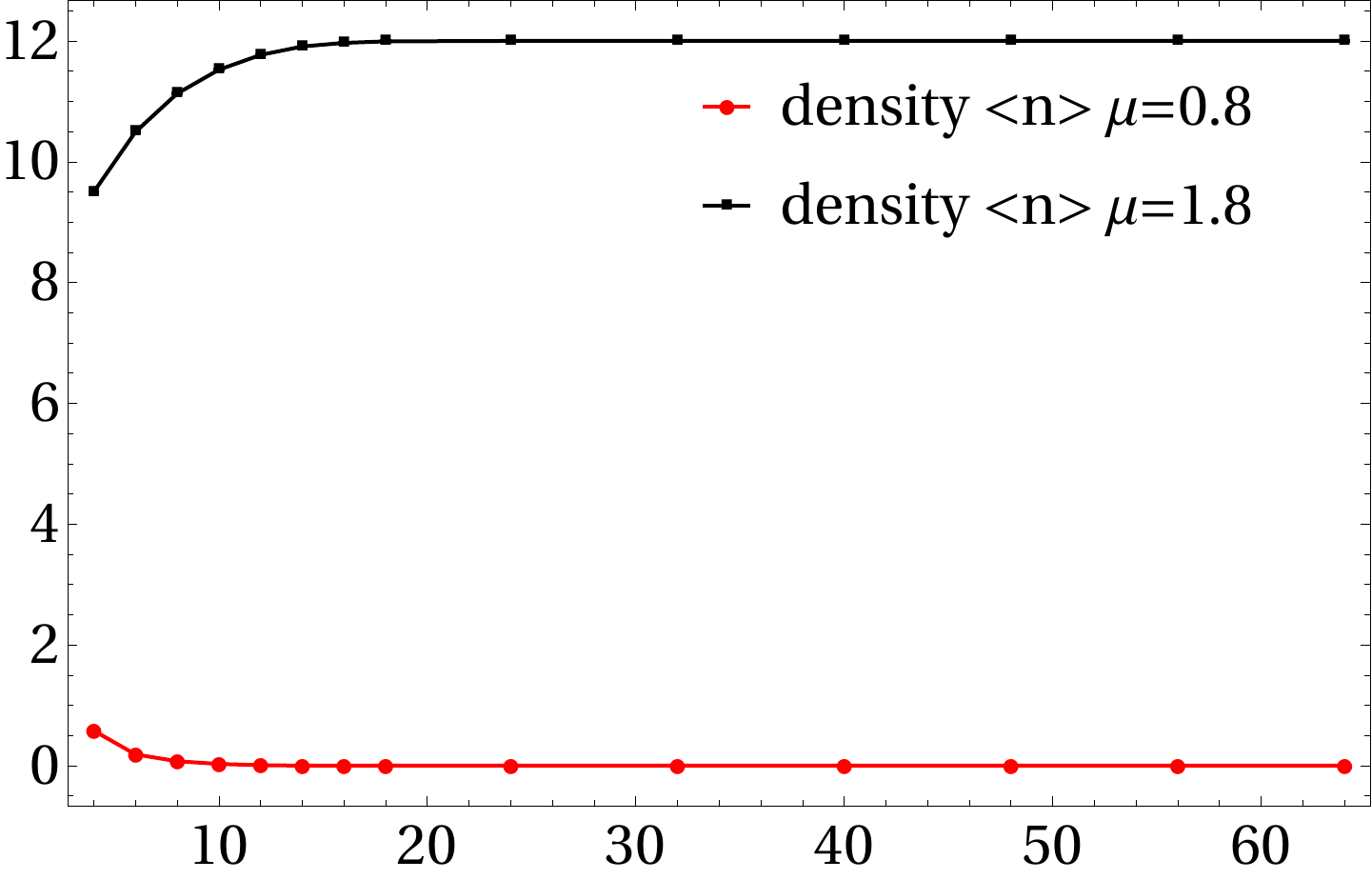}
\caption{QCD in $\kappa$ expansion, $\beta=6, \kappa=0.12$.
{\it Top}:  $\mu$-dependence of the Polyakov loop ({\it left})
and of the  density ({\it right}) at fixed $N_{\tau}=16$. {\it Bottom}: 
 $N_{\tau}$-dependence of the Polyakov loop ({\it left}) and  
 of the density at fixed $\mu$ ({\it right}).
}
\label{fig:lowT}\end{figure}

\section{Further prospects: $SU(2)$  real time simulations}\label{further}

Revisiting direct simulations of the Minkowski action 
in $SU(2)$ \cite{Berges:2007nr},
\cite{Berges:2006xc}, 
the use of gauge cooling  \cite{Seiler:2012wz} and dynamical 
stabilisation \cite{Attanasio:2016mhc} show promising improvements. 
Various integration contours are possible \cite{Berges:2006xc}, \cite{Alexandru:2016gsd},
 we found as optimal 
an isosceles triangle with base of extension $\tau$
along the imaginary axis and of extension $t_0$ in the real time direction,
 discretised in $N_t$ points. 

The left plot of Fig. \ref{fig:SU2} depicts the maximum possible 
real time extent $t_0$ of 
 the contour for fixed $\tau=16$ and $N_t=16$. Here gauge fixing fails 
but gauge cooling and dynamical stabilisation allow for a non-zero 
 real-time extent in an interesting $\beta$-region. Note that for small
  $N_t$ there can be deviations in the expectation value of the spatial 
  plaquette between the euclidean and the real-time contour. This difference
   however becomes smaller as one increases $N_t$. The right plot 
   depicts the difference as a function of $N_t$ at fixed $t_0$.
            . 
The discretisation of $t_0$, $N_t$, is relevant. 
Larger $\beta$ allow larger real time extents. 
The maximum extent of $t_0$ strongly depends on $\tau$. 
For smaller $\tau$  larger ratios $t_0 / \tau$ 
are possible. 
\begin{figure}[thb] 
  \centering
  \includegraphics[width=6.2cm,clip]{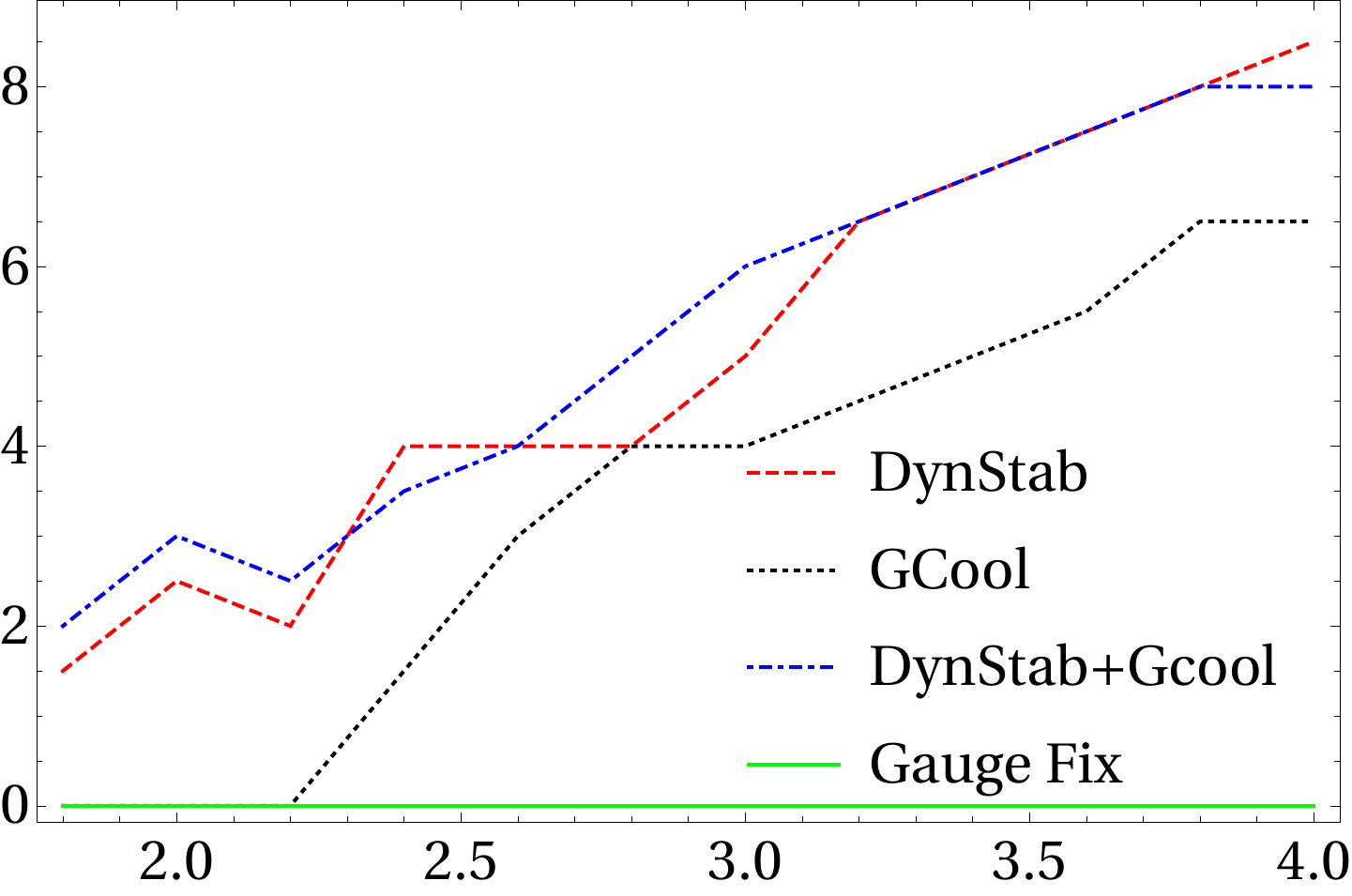}
  \includegraphics[width=7.cm,clip]{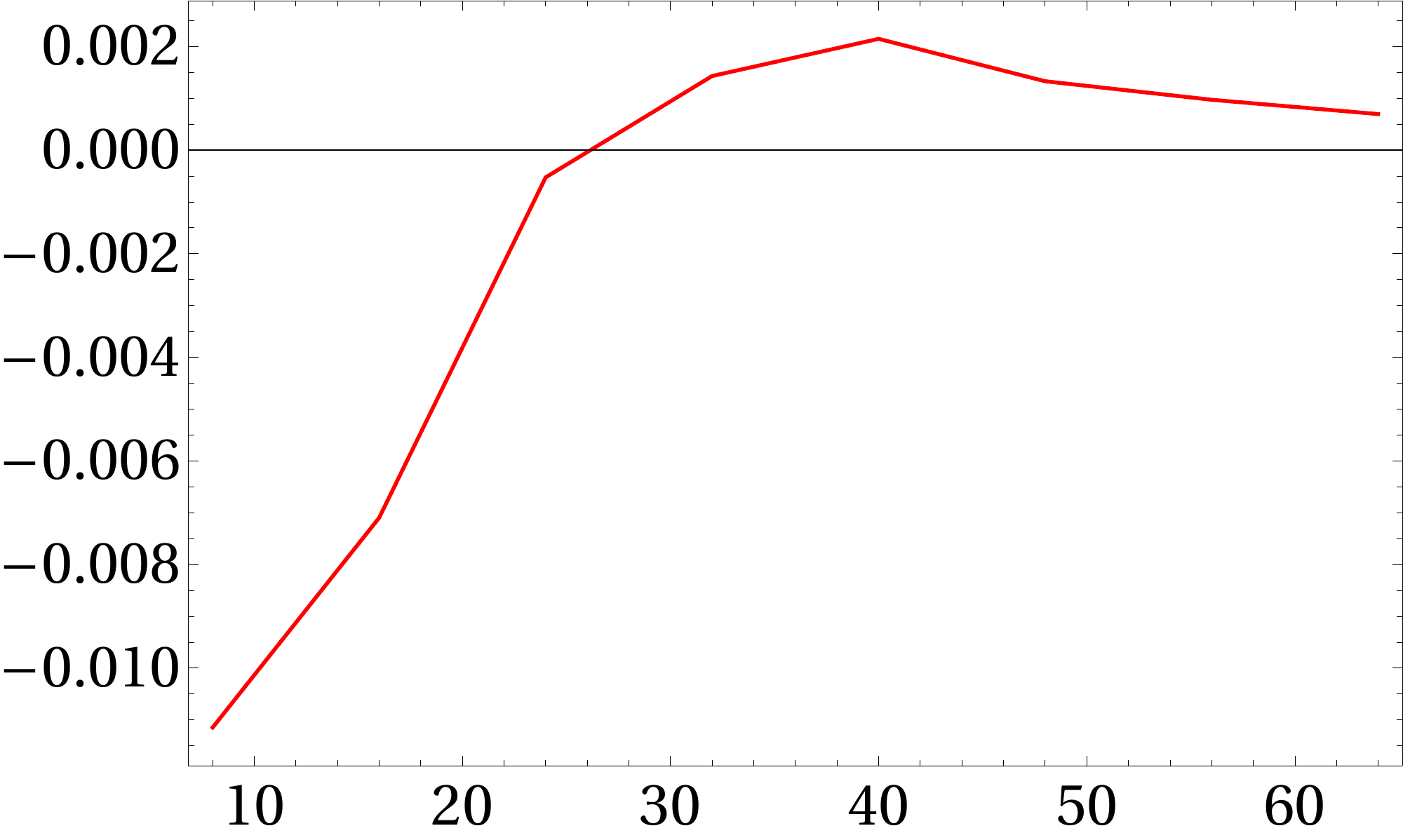}
  \caption{$SU(2)$ real time evolution. {\it Left} Maximal possible 
  real time extent $t_0$ vs. $\beta$ for 
$N_s=8$, $N_t=16$ and inverse temperature $\tau=16$. {\it Right}: 
Difference of the spatial plaquette between the euclidean and the symmetric 
triangle contour with real time extent $t_0=3$ as a function of $N_t$ at 
$\beta=3$, $N_s=16$.}
  \label{fig:SU2}
\end{figure}
\begin{figure}[h]
\includegraphics[width=0.5\textwidth]{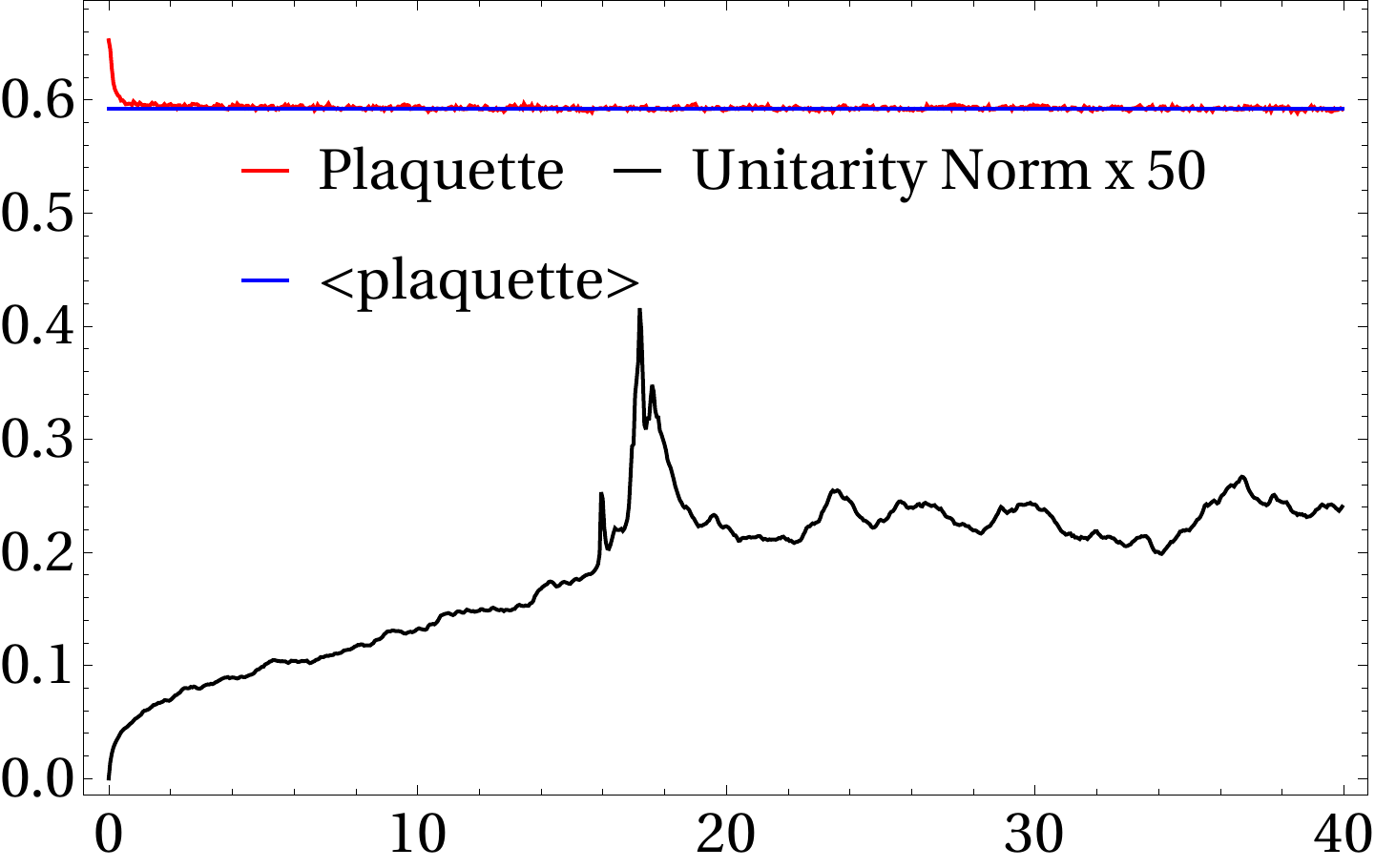}
\includegraphics[width=0.5\textwidth]{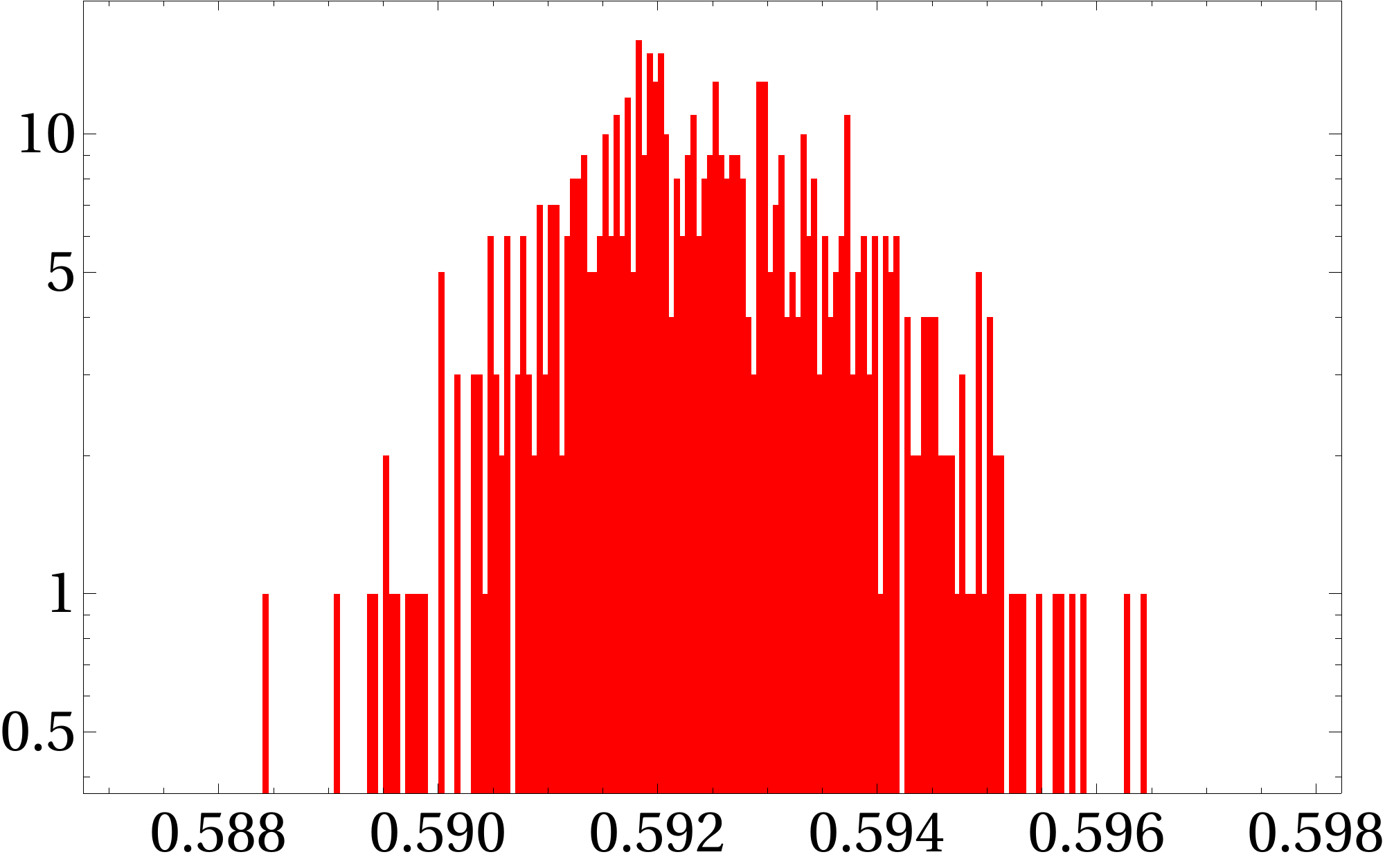}
\includegraphics[width=0.5\textwidth]{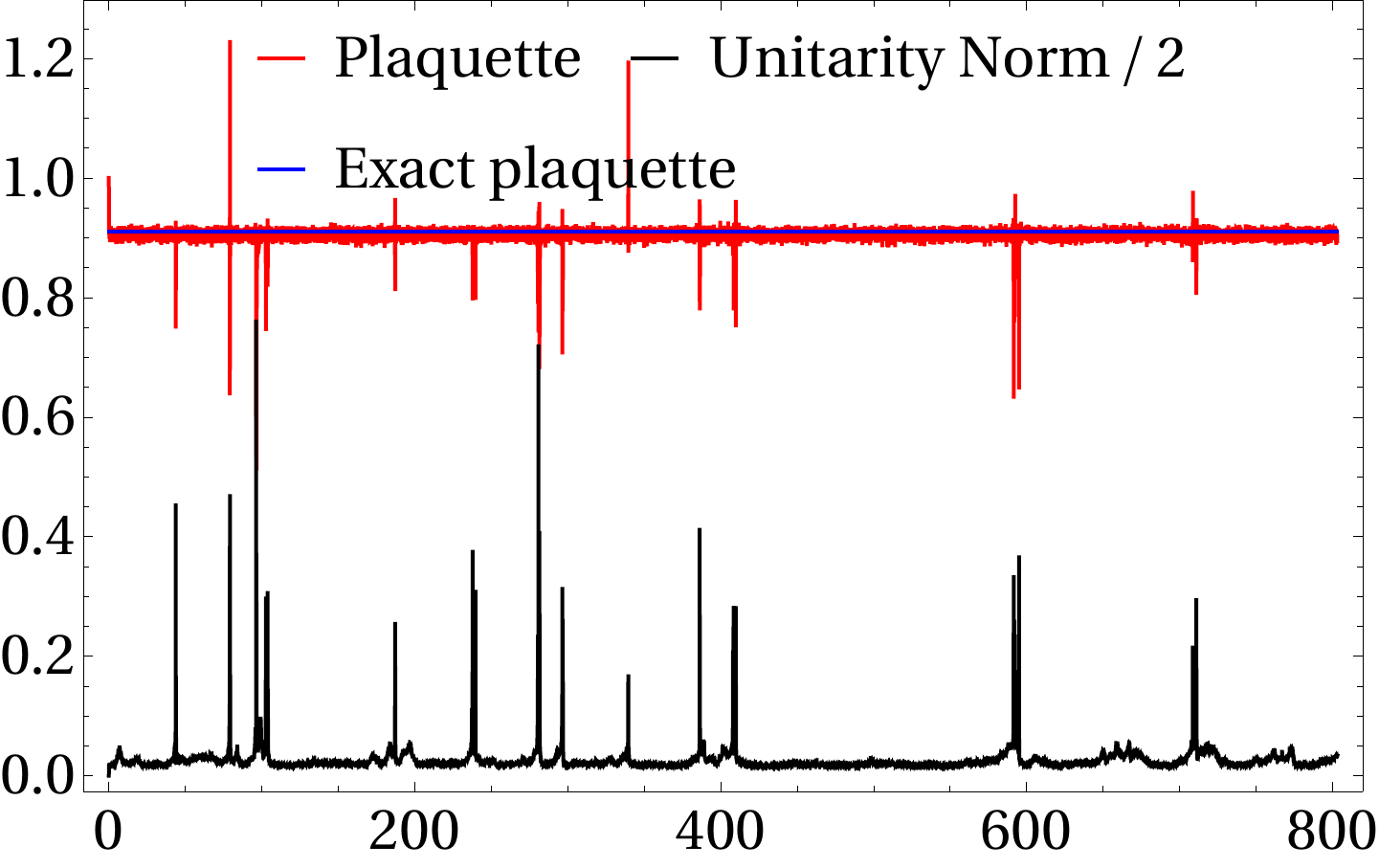}
\includegraphics[width=0.5\textwidth]{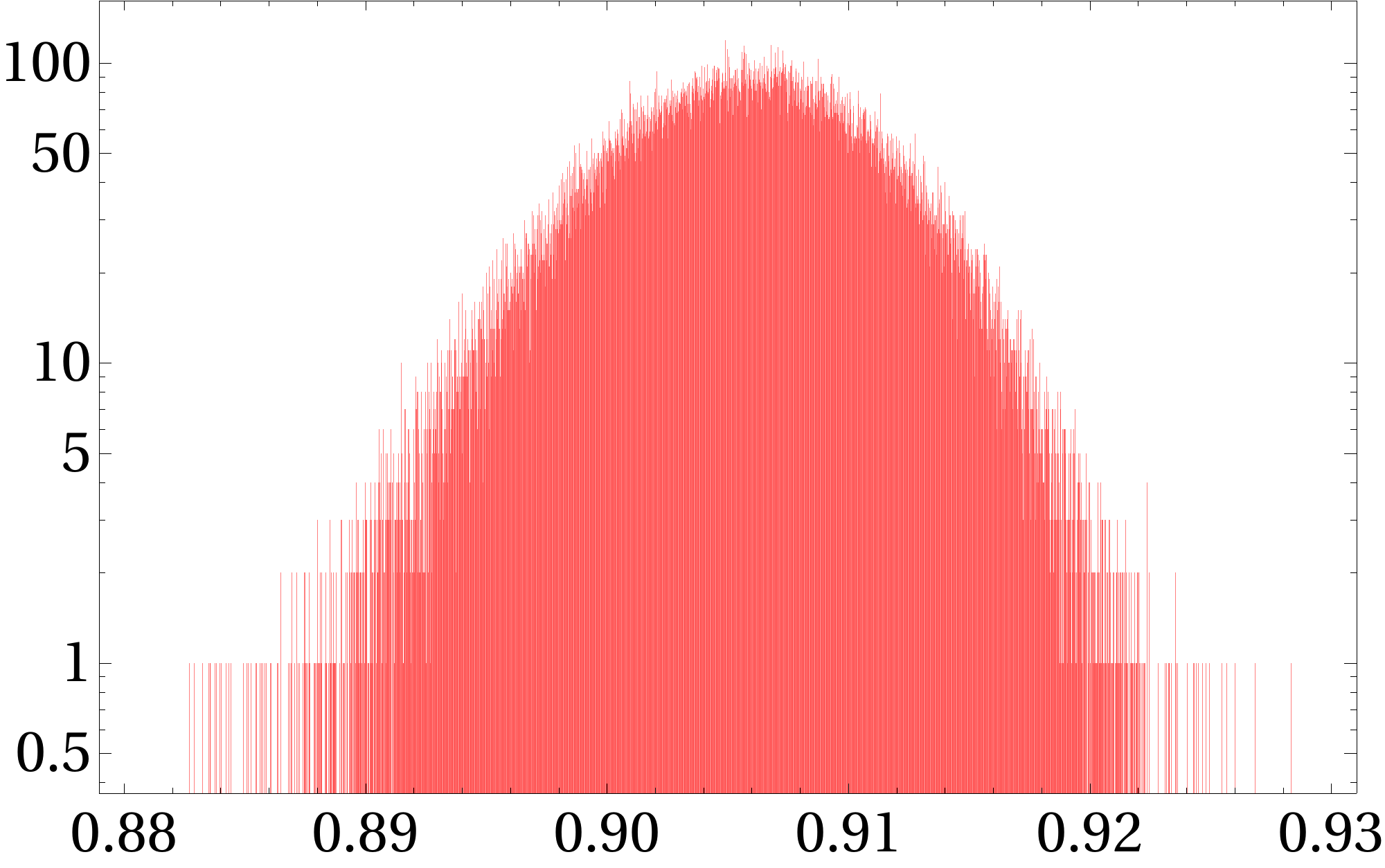}
\caption{Cooling effectiveness. {\it Top}: QCD in $\kappa$ expansion, 
$\beta=6, \, \kappa=0.12,\, N_{\tau} =16$, \,  Langevin evolution 
with cooling (12 sweeps per dynamical sweep) of ${\cal N}_U$ 
and the plaquette ({\it left}) 
and histogram of the latter  using the regions
of bounded ${\cal N}_U$ ({\it right}). {\it Bottom}: $SU(2)$ real time on the 
isoscel triangle, $N_s=4$, $N_t=8$, $\beta=16$,\,  
Langevin evolution with cooling (2  sweeps) of 
the unitarity norm and the spatial plaquette 
({\it left})  and histogram of the plaquette using 
${\cal N}_U < 0.07$ ({\it right}). Before discarding the 
peaks in ${\cal N}_U$ the histograms had pronounced  skirts.
}
\label{cool}
\end{figure}

\section{Conclusions}\label{conc}

{\em Gauge cooling} \cite{Seiler:2012wz}
(and its further developments) appear essential for
controlling the 
 behaviour of the CL distributions  
both in the far non-compact direction 
 and near the poles
 \cite{Aarts:2017vrv}, see also  \cite{review}.
 The efficiency of the CL method for lattice gauge theories
 therefore strongly
depends on the effectiveness of cooling. The latter can
be judged from the unitarity norm ${\cal N}_U$ which is expected
to stabilise at some value fixed by the dynamics. 
In the simulation numerical errors
may eventually lead to a divergence of ${\cal N}_U$, 
but as long as ${\cal N}_U$
stays bounded we can sample reliable results. 

 The evolution of the  ${\cal N}_U$
 under cooling permits to judge the 
 efficiency of the latter and therefore the reliability of the results
  - see Fig. \ref{cool}.
 Isolated peaks do not spoil the simulation as long as 
 there are sufficiently long intervals of small unitarity norm. 
 Removing the occasional
 peaks and selecting the intervals of small
 ${\cal N}_U$ leads then to well behaved distributions and
 correct convergence of the simulation (remember that cooling is a
gauge transformation).  
For tests of other cooling procedures see, e.g. \cite{Nagata:2016vkn}, 
\cite{Nagata:2016mmh}, \cite{Attanasio:2016mhc}.  

For 
QCD lattice calculations at $\mu >0$
our results in a 10-th order $\kappa$ expansion show that 
simulations in the confinement 
and deconfining phases are possible
for $\beta$ large enough. Work in progress
concerns resolving the phase transition region. Low
temperatures require large lattices in order
to ensure a good aspect ratio. 
A lesson from the study of simple models
is to check the appearence
of domains with $\Rep D <0$, 
drop their contributions and  estimate a possible
systematic error. As shown in \cite{Aarts:2017vrv}
the signal is already visible in the observables and can
thus be easily monitored.

For SU(2) real time evolution cooling and dynamical stabilisation 
are shown to improve  the simulations  also for smaller $\beta$, future work is
however still necessary to achieve realistic scenarios.


\bibliography{lattice2017}

\begin{thebibliography}{99}
{\small
%
   \bibitem{review} G. Aarts, L. Bongiovanni,   E. Seiler, 
  D. Sexty, I.-O. Stamatescu,
  Eur.\ Phys.\ J.\ A {\bf 49} (2013) 89
  [arXiv:1303.6425 [hep-lat]]
%
%
\bibitem{Seiler:thiscontrib65} E.~Seiler, these proceedings.
\bibitem{Aarts:2009uq}
  G.~Aarts, E.~Seiler and I.~O.~Stamatescu,
  Phys.\ Rev.\ D {\bf 81} (2010) 054508
  [arXiv:0912.3360 [hep-lat]].
\bibitem{Aarts:2011ax}						       
  G.~Aarts, F.~A.~James, E.~Seiler and I.~O.~Stamatescu,	        
  Eur.\ Phys.\ J.\ C {\bf 71} (2011) 1756			        
  [arXiv:1101.3270 [hep-lat]].					        
\bibitem{Aarts:2017vrv}
  G.~Aarts, E.~Seiler, D.~Sexty and I.-O.~Stamatescu,
  JHEP {\bf 1705} (2017) 044
  [arXiv:1701.02322 [hep-lat]].
%
\bibitem{Aarts:2012ft}
G.~Aarts, F.~A. James, J.~M. Pawlowski, E.~Seiler, D.~Sexty and I.-O.
  Stamatescu, 
  JHEP {\bf 03} (2013) 073, [arxiv: 1212.5231 [hep-lat]].
%
\bibitem{Seiler:2012wz}
  E.~Seiler, D.~Sexty and I.-O.~Stamatescu,
  Phys.\ Lett.\ B {\bf 723} (2013) 213,
[arXiv:1211.3709 [hep-lat]].
%
\bibitem{Nagata:2016vkn}
  K.~Nagata, J.~Nishimura and S.~Shimasaki,
    Phys.\ Rev.\ D {\bf 94} (2016) no.11,  114515
      [arXiv:1606.07627 [hep-lat]].  
%
\bibitem{Nagata:2016mmh}
  K.~Nagata, H.~Matsufuru, J.~Nishimura and S.~Shimasaki,
  PoS LATTICE {\bf 2016} (2016) 067
  [arXiv:1611.08077 [hep-lat]].
\bibitem{Attanasio:2016mhc}
F.~Attanasio and B.~J{\"a}ger, 
  PoS {\bf LATTICE2016} (2016) 053,
  [arxiv:1610.09298[hep-lat]].
\bibitem{Stamatescu:2016dwx} 
  I.-O.~Stamatescu and E.~Seiler,
  PoS LATTICE {\bf 2016}, 357 (2016)
  [arXiv:1611.00620 [hep-lat]].
\bibitem{Bender:1992gn} 
  I.~Bender, T.~Hashimoto, F.~Karsch, V.~Linke, A.~Nakamura, M.~Plewnia, I.-O.~Stamatescu and W.~Wetzel,
  Nucl.\ Phys.\ Proc.\ Suppl.\  {\bf 26}, 323 (1992).
\bibitem{DePietri:2007ak}
  R.~De Pietri, A.~Feo, E.~Seiler and I.-O.~Stamatescu,
  Phys.\ Rev.\ D {\bf 76} (2007) 114501
  [arXiv:0705.3420 [hep-lat]].
%
\bibitem{Aarts:2016qrv}
  G.~Aarts, F.~Attanasio, B.~Jäger and D.~Sexty,
  JHEP {\bf 1609} (2016) 087
  [arXiv:1606.05561 [hep-lat]].
\bibitem{Aarts:2014bwa} 
  G.~Aarts, E.~Seiler, D.~Sexty and I.-O.~Stamatescu,
  Phys.\ Rev.\ D {\bf 90}, 114505 (2014),
  [arXiv:1408.3770 [hep-lat]].
\bibitem{Fromm:2011qi}
  M.~Fromm, J.~Langelage, S.~Lottini and O.~Philipsen,
  JHEP {\bf 1201} (2012) 042
  [arXiv:1111.4953 [hep-lat]].
\bibitem{Bloch:2017jzi}
J.~Bloch and O.~Schenk,
  [arXiv:1707.08874 [hep-lat]].
\bibitem{Sinclair:2016nbg} 
  D.~K.~Sinclair and J.~B.~Kogut,
  PoS LATTICE {\bf 2016}, 026 (2016),
  [arXiv:1611.02312 [hep-lat]].
\bibitem{Schmalzbauer:2016pbg}
  S.~Schmalzbauer and J.~Bloch,
  PoS LATTICE {\bf 2016} (2016) 362
  [arXiv:1611.00702 [hep-lat]].
\bibitem{Sexty:2013ica}
D.~Sexty,
  Phys.\ Lett.\ B {\bf 729} (2014) 108
 [arXiv:1307.7748 [hep-lat]].
\bibitem{Berges:2007nr}
  J.~Berges and D.~Sexty,
  Nucl.\ Phys.\ B {\bf 799} (2008) 306
  [arXiv:0708.0779 [hep-lat]].
\bibitem{Berges:2006xc}
  J.~Berges, S.~Borsanyi, D.~Sexty and I.-O.~Stamatescu,
  Phys.\ Rev.\ D {\bf 75} (2007) 045007
  [hep-lat/0609058].
  \bibitem{Alexandru:2016gsd}
  A.~Alexandru, G.~Basar, P.~F.~Bedaque, S.~Vartak and N.~C.~Warrington,
  Phys.\ Rev.\ Lett.\  {\bf 117} (2016)   081602
  [arXiv:1605.08040 [hep-lat]].
}

\end{thebibliography}

\end{document}